\documentclass{jors}
\usepackage{graphicx}
\usepackage{listings}
\usepackage{natbib}

\definecolor{mygray}{gray}{0.6}

\begin{document}

\section*{(1) Overview}
\thispagestyle{empty} 

\vspace{0.5cm}

\section*{Title}

AskCI Server: Collaborative and version controlled knowledge base

\section*{Paper Authors}
1. Sochat, Vanessa;

\section*{Paper Author Roles and Affiliations}
1. Stanford University Research Computing

\section*{Abstract}

AskCI server is a collaborative, open source documentation server that uses GitHub for automation and version control of shared knowledge. A programmatic application programming interface, friendly user interface, and organization of concepts into questions makes it versatile as a support or collaborative knowledge base.

\section*{Keywords}

documentation; version control; collaborative; articles; knowledge; automated; docker;

\subsection*{Introduction}

Institutional knowledge about resources at academic centers pertaining to high performance computing,
research software, and other ideas or concepts centered around cyberinfrastructure have historically been
siloed. A research computing center is likely to provide documentation specific to their resources, or 
a small technology group to provide support catered to their user base. Although there are benefits to
maintaining isolated and specific documentation, the drawbacks for the community can be substantial.
A researcher using the documentation typically experiences a large set of technologies in synchrony,
and would benefit from a more global context that captures knowledge across support groups.
 It is common to see redundancy in academic documentation bases, as many research groups or resources tend to use
similar technologies. By way of having small teams work on these documentation bases, the potential for
stale or outdated material is large. The negative consequences largely fall on the user, as they must
sift through several online resources to put together a current, holistic picture of a concept or tool.
The negative consequences also fall on the centers, as they struggle to support the user bases with
smaller teams, and have limited time to allocate to support and documentation. In that support by way
of tickets or in-person help is essential, documentation usually falls to the wayside.
Badly needed is a collaborative solution to take the burden off of any individual center or group, and
one that moves knowledge into the open source domain.

\section*{Concepts}

AskCI Server \cite{askci-server} is a version controlled, collaborative knowledge and support server. It
introduces the powerful concept of open source knowledge, or bringing best practices from open source development
to the maintenance of knowledge. This means that knowledge is version controlled, worked on collaboratively
across institutions, available programmatically, and syndicated in a central location. 
Specifically, AskCI Server is defined by the following concepts:
\newline

\thispagestyle{empty} 

\begin{enumerate}
 \item Articles: Topics or concepts that a user might want to ask a question about. On a high level, it's a piece of knowledge that can be collaboratively worked on. On a functional level, an article corresponds to a single GitHub repository based on a template specification that allows for interaction with the server \cite{tech-spec}.
\item Questions and Examples: Embedded inquiries or code snippets in an article that are indexed and searchable. A user can search and find a specific question, and then be taken immediately to the location in an article's text where the answer resides.
 \item Reviews: Are submit by an authenticated user on the site, and map to pull requests (PRs) on GitHub \cite{github-pr}. In the same way that pull requests can be used to discuss changes to software, the environment is equally friendly to discuss changes to knowledge articles. When a pull request is merged, the AskCI Server is notified and the knowledge article is updated.
 \item User: Can be a visitor (non-authenticated), an editor or reviewer (authenticated but without ownership of knowledge repositories) or an owner (authenticated with ownership). Visitors can browse content, editors and reviewers can update or ask new questions, and owners can do all of the above plus serve as maintainers for the knowledge repositories.
\end{enumerate}

In practice, this means that content is created, worked on, and updated on GitHub \cite{github}, and each article (repository) provides knowledge, examples, and interesting content links (references, tutorials, etc.) pertaining to a single concept or idea, akin to Wikipedia \cite{wikipedia}. Interactions between GitHub and the AskCI Server are automated via webhooks and GitHub workflows. Since questions are embedded in articles and then indexed by the server, a user is allowed to ask a question via the interface or a connected tool to easily find an answer or code snippet example.

\section*{The AskCI Community}

AskCI Server is branded alongside the discourse server ask.ci \cite{askci} as the two can serve different needs for the same community. However, the two serve distinct use cases. While the discourse ask.ci \cite{askci} is akin to a discussion based forum where concepts might appear on many topics scattered across the site, AskCI Server \cite{askci-server} provides a single article for each concept. \newline

Topics on ask.ci, by way of being organized in a question and answer format [@discourse] do not represent holistic knowledge, but rather individual questions that a user would still need to browse through to form some cohesive understanding. While this structure is useful in that it allows individual centers to create support forums for their users (categories called "locales"), it isn't a design
well suited to creating holistic knowledge. AskCI Server, on the other hand, acts more like a wiki with articles for meta- or super- concepts such as a container technologies, job managers, or software projects. If a user is searching for the answer to a question about a particular topic, he or she can be confident that if the question has been answered, it will be found on the topic's page. If it's not answered, the user can easily ask the question for the maintainers of the term to answer. In that the discourse ask.ci is scattered with snippets of high quality knowledge that might be organized to a concept, there is a logical progression for this content from the ask.ci discourse site to wind up on the AskCI Server. A webhook can be set up to enable this process. 

AskCI Server can support the following properties and use cases:

\thispagestyle{empty} 
\section*{Version Control}

In that each article, a concept or idea, is maintained under version control \cite{github}, this means that 
a history is kept for all changes over time pertaining to its development. If an AskCI Server itself
were to go away for any reason, the repositories and knowledge would live on, and could easily be imported
into another deployment.

\section*{Asking Questions}

A user browsing the site is first presented with the most recent articles and associated questions (Figure 1),
and the site is organized to make it easy to browse articles, reviews under progress, or embedded questions and examples.

\begin{figure}[h]
\caption{Questions are Presented on the main page}
\centering
\includegraphics[width=1.0\textwidth]{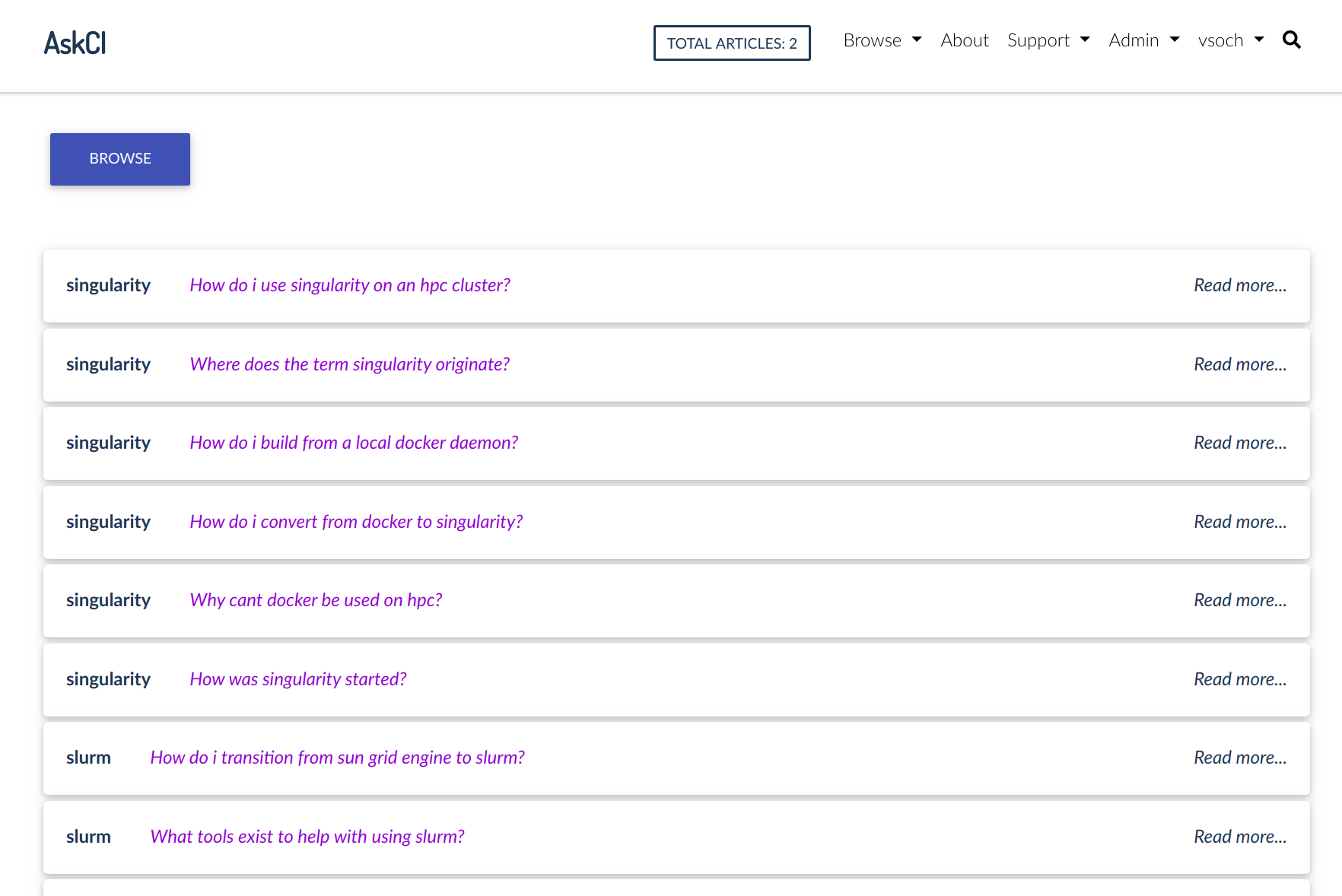}
\end{figure}

The user is also able to easily search all content across the site, which is also made available
via the AskCI Server Application Programming Interface (API). Command line tools to streamline
asking and answering questions can be developed using this API, and questions are linked
to specific locations in the text. This is made possible by way of span tags. 
While not visible in the rendered markdown, questions and examples are embedded in the content
by way of these tags. A span for a question or example can appear anywhere in the text, and marks the location where
the content beings. An example question might look like this:

\begin{lstlisting}[language=html]
<span id="question-how-did-hpc-originate"></span>
\end{lstlisting}

An example would appear before a code block.  Clicking on a question or example from an article's page or otherwise navigating to the span highlights the location for the user. Both of these structures can be inserted easily in the markdown editor on an article's page, and content is tested automatically. This automation is discussed next.

\thispagestyle{empty} 
\section*{Automated Workflows}

Each term repository is based on a common template \cite{tech-spec} that comes with GitHub Workflows \cite{github-workflows} that
can fully automate management of the term. These workflows include:

\begin{enumerate}
\item Testing: Testing of the term content, or the README.md maintained in the repository, comes down to parsing the text for correctly structured example and question spans. The content is tested in the repository, and submissions are also tested on the server. A user that is requesting review is not be able to submit until validation passes.
\item Request for Review: A request for review is done by way of a dispatch event \cite{github-dispatch} that is identified based on a client metadata field, "request-review." When a user edits content and submits it for review, the dispatch event will receive the updated content, open up a new branch on the repository with the content, and open a pull request. The submitting author is notified on the pull request to allow for further discussion, and the pull request is linked from the term interface and site for others to see and give feedback on.
\item Template Update: An update to the template would be highly challenging if the template was used across, say, hundreds of repositories. To support ease of updating, a dispatch event identified by a client metadata field "update-template" can be triggered by an administrator of the server to update all or a subset of templates. An update comes down to cloning the upstream template, and updating the hidden .github folder scripts that drive the application. A pull request is opened for the term maintainers to review the changes.
\end{enumerate}

\section*{Webhooks}

Along with workflows, webhooks are essential for keeping the latest content on GitHub in sync with the server. Webhooks are created automatically when a new term is added, which also coincides with creating a newly named copy of a term template. Webhooks include:

\begin{enumerate}
\item Push or Deployment: Any update of the term content in its README.md that is pushed to the master branch notifies the server. The server validates the webhook, and then updates content on the server from the repository. 

\item Pull Request: As pull requests are linked from the server to solicit additional review, it's essential that any changes in status are also kept in sync. Whenever a pull request status is changed, the server is notified, and acts accordingly. For example, a pull request closing without a merge would result in a status of "reject" for the review, while a pull request with a merged time stamp would indicate that it was accepted and closed.
\item Repository: In the case that any information about a repository changes, including the name, owner, or metadata, a webhooks is triggered to update the server. If the name is changed to be outside of the AskCI Server namespace (in the format `askci-term-<term>`) or if any subsequent actions on the repository are not successful, then the repository is marked as archived on the server. Archived means that further updates will not happen unless the permissions or metadata issues are resolved. This repository webhook will also look for updated topics \cite{github-topics} on the repository, which are added as tags to the article.
\end{enumerate}

\thispagestyle{empty} 
\section*{Notifications}

For all of the above cases, in that a request for review, template update, or repository archive occurs,
owners are notified by way of an email powered by the SendGrid API \cite{sendgrid}. Users that contribute
content to repositories are also subscribed to receive further notifications, a status that they can
easily remove if desired.

\section*{Contributing Content}

There are several ways for which incentive exists for generating updated content.

\begin{enumerate}
\item Contributor: Any authenticated user, even with minimal GitHub permissions, can edit an article, and then submit the changes to the open source knowledge maintainer for review (Figure 2). This is likely to happen if a user is browsing the site and sees content that he or she would like to update.
\item Asking Questions: If a browsing user cannot find an answer to a question on any particular topic, he or she can easily click on "Ask a Question" and input the question content. The question is submit as a GitHub issue to the repository associated with the term, and brought to the attention of the open source knowledge maintainer(s). This question asking most likely will happen in a web interface, but could also programmatically happen by way of the application programming interface.
\item Discourse Webhook: An external webhook can be configured to ping the AskCI Server when there is a post on an external discourse board. In the case of ask.ci \cite{askci}, a post would come down to a user asking a specific question about some cyberinfrastructure topic. When the post is added to the ask.ci discourse server, it pings the webhook with it's content, and the AskCI Server parses the content, matches terms to existing terms on the server, and if a match is found, opens up an issue on GitHub to alert any maintainers that new knowledge exists that might be integrated into the article.
\end{enumerate}

\begin{figure}[h]
\caption{Markdown Editing Interface for topic "Singularity"}
\centering
\includegraphics[width=1.0\textwidth]{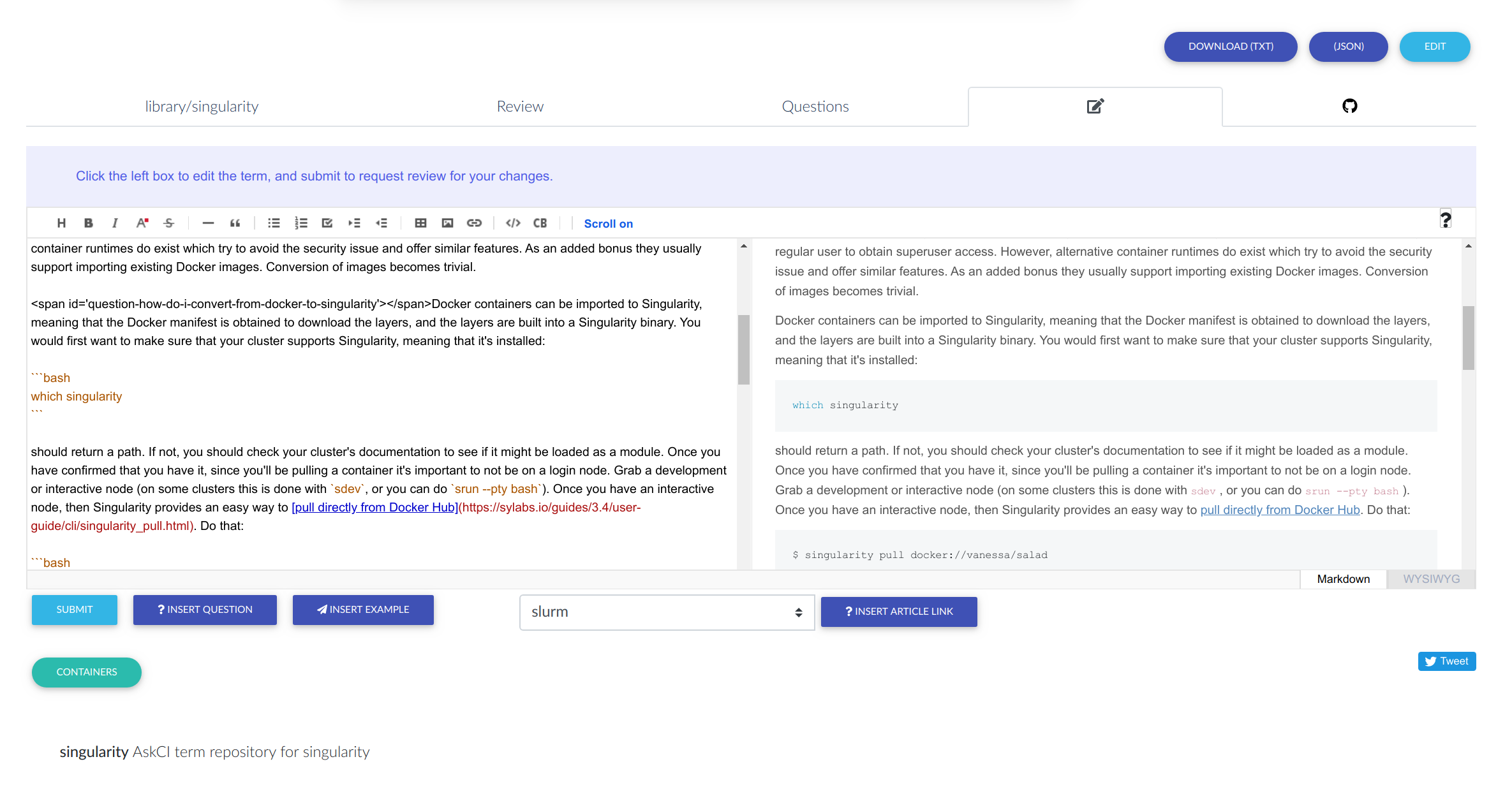}
\end{figure}

The webhook framework is generalized, so any other knowledge resource that supports sending webhooks could be integrated.

\section*{Implementation and architecture}
\thispagestyle{empty} 

\paragraph{Authentication}
There are four levels of roles provided by AskCI Server, three of which are available to the larger community.

\begin{enumerate}
\item Viewer: Any unauthenticated user is allowed to browse knowledge on the site, including articles, questions, reviews, and examples.
\item Editor: A role associated with allowing read-only access for the minimum set of public information provided by GitHub OAuth2 \cite{github-oauth2}. An editor is able to interactively edit the article content in the interface, and then submit the changes as a request for review.
\item Owner: A role associated with read and write permissions required to copy template repositories and create webhooks on the user's behalf. An owner can easily generate new knowledge repositories for terms or concepts that do not exist yet. An owner is then responsible for being one of the maintainers of the term on GitHub, and will receive notification for reviews or new questions. If a repository is generated in a GitHub organization, this responsibility can be shared by a group of individuals with shared expertise, and this is the recommended approach.
 \item Admin: A site administrator is considered a staff or superuser of the application, meaning that she has full access to manage the site. Typically an admin would be responsible for responding to issues with respect to the site, updating templates when necessary, or setting up webhooks.
 \end{enumerate}

The flexibility with respect to roles allows for a user of the server to participate at whatever level is comfortable for him or her. Some will want to read passively, others will want to contribute content without taking on responsibility, and a third group will want to serve as open source knowledge maintainers.

\paragraph{Containers}

The AskCI Server is made possible by way of several Docker \cite{docker} containers, including:

\begin{enumerate}
\item askci\_base: A uwsgi \cite{uwsgi} container that runs the main Django \cite{django} application
 \item askci\_worker: A django\_rq \cite{django-rq} worker that can run asynchronous tasks
 \item askci\_scheduler: A scheduler for the worker
 \item askci\_redis: The redis database for the scheduler
 \item askci\_postgres: A database for the application
\end{enumerate}

These containers are deployed locally or in production by way of docker compose \cite{docker-compose}. A shell script, askci.sh is provided alongside the application for easier interaction to deploy or manage a development or production interface.
For production, it's suggested to use a database outside of a container, and essential to be deployed with https.

\thispagestyle{empty} 
\section*{(2) Availability}
\vspace{0.5cm}
\section*{Operating system}
Linux

\section*{Programming language}
Docker and docker compose

\section*{Additional system requirements}
Ports 80 and 443

\section*{Dependencies}
Docker and docker-compose

\section*{Software location:}

\begin{description}[noitemsep,topsep=0pt]
	\item[Name:] AskCI Documentation Server v1.0.0
	\item[Persistent identifier:] https://zenodo.org/badge/latestdoi/223678399
	\item[License:] Mozilla Public License 2.0
	\item[Publisher:]  Vanessa Sochat
	\item[Version published:] 1.0.0
	\item[Date published:] 01/12/20
\end{description}

{\bf Code repository} GitHub

\begin{description}[noitemsep,topsep=0pt]
	\item[Name:] vsoch/askci
	\item[Persistent identifier:] https://github.com/vsoch/askci
	\item[License:] Mozilla Public License 2.0
	\item[Date published:] 11/23/19
\end{description}

\thispagestyle{empty} 
\section*{Language}
Python, Docker

\section*{(3) Reuse potential}

While this particular example has scoped the AskCI Server to be about research computing and technology support, and is
branded alongside the discourse ask.ci \cite{askci}, the AskCI Server is not limited to this community or use case. 
An AskCI Server can easily be branded to support any kind of knowledge that can be maintained on GitHub. The setup steps walk the user through the server naming, and this can be easily changed. While the main
README.md was selected for this early template so that the main repository also renders the knowledge content, for a software repository
that uses the main README.md for other purposes, a template could be customized to render one or more different files.
An AskCI Server could easily be branded to serve metadata about software, medical or biological knowledge, or specbifications.
Further, by way of the application programming interface, an AskCI Server can be integrated with an external
ticketing system, support queue, or similar. \newline

The beauty of the design of this documentation base is that knowledge repositories, even if disconnected from
a central server, can be imported easily into a new server. For this reason, articles that are written during the development
of an AskCI Server will integrate easily into a production server. For more details about usage and development, screen shots, and
links, see the official documentation at https://vsoch.github.io/askci \cite{askci-server}.

\section*{Acknowledgements}

Author V.Sochat would like to thank Katia Bulekova for her detailed testing of the interface, and the AskCI community for general support.

\section*{Competing interests}
The authors declare that they have no competing interests.

\newpage
\thispagestyle{empty} 
\bibliography{paper}
\bibliographystyle{bib.bst}

\vspace{2cm}

\end{document}